\documentstyle[twoside,fleqn,espcrc2]{article}
\include{epsf}
\newcommand{\AmS}{{\protect\the\textfont2
  A\kern-.1667em\lower.5ex\hbox{M}\kern-.125emS}}

\title{Spectroscopy of Hadrons with $b$ Quarks from Lattice NRQCD}

\author{A.~Ali Khan\address{Physics Department, The Ohio State University,
Columbus, OH 43210, USA}\thanks{In collaboration with T.~Bhattacharya,
S.~Collins, C.~T.~H.~Davies, R.~Gupta, C.~Morningstar, J.~Shigemitsu
and J.~Sloan. 
Present address: Center for Computational Physics, University of Tsukuba,
Tsukuba, Ibaraki 305, Japan. }%
}
\begin{document}

\begin{abstract}
Preliminary results from an extensive lattice calculation of the $B$, 
$B_c$, and $\Upsilon$
spectrum at quenched $\beta = 6.0$ are presented. The study includes 
radially and orbitally excited mesons, and baryons containing $b$ quarks.
The $b$ quarks are formulated using NRQCD; for light and $c$ quarks, a
tadpole-improved clover action is used.
\end{abstract}

\maketitle

\section{INTRODUCTION}
Experimental data on excited $B$ mesons and $b$  baryons
have begun to emerge just recently. One hopes that in the next years, these
states will be firmly established and accurate data on their masses become
available. For the $B_c$, experimental results are so far restricted to
preliminary values for the ground state mass. 

It is therefore of great interest to obtain predictions for the $B$ and $B_c$
spectrum from lattice QCD. This talk is a status report on a calculation using 
NRQCD for the $b$ quarks, and a tree level tadpole-improved clover action for
the light and charm quarks. The simulation was done quenched at $\beta = 6.0$
on a lattice volume $16^3 \times 48$. The NRQCD action and the simulation
parameters for the heavy-light physics are described in
Ref.~\cite{fBpaper}. Our two $\kappa$ values around the charm are 
$0.119$ and $0.126$.
All the results presented in this article are preliminary.

\section{THE HEAVY-LIGHT SPECTRUM}
For the $B$ spectrum results 
presented here we fix the lattice spacing from $M_\rho$ and obtain $a^{-1} =
1.92(7)$ GeV. Our determination of the averaged $u$ and $d$, and  strange,
quark masses is described in Ref.~\cite{fBpaper}. 
%We calculate the heavy-light meson mass using the relation,
%\begin{equation}
%M = E_{\rm sim} + \Delta_{\rm NRQCD},
%\end{equation}
%where $E_{\rm sim}$ is the exponential falloff of the correlator, and 
%$\Delta_{NRQCD}$ has been obtained from the heavy-heavy dispersion
%relation:
%\begin{equation}
%\Delta_{NRQCD} = 1/2(M_{\rm kin}(HH) - E_{\rm sim}(HH).
%\end{equation}
For our heavy-light meson mass we use the definition
$M^\prime$ of Ref.~\cite{fBpaper}. We fix the $b$ quark mass in a slightly
different way from that in Ref.~\cite{fBpaper}. Instead of using the pseudoscalar
meson mass, we set the
spin-averaged meson mass to the physical value: $1/4(M_B + 3M_{B^\ast}) = 5313$ 
MeV. 
\subsection{Mesons}
\begin{figure}[t]
%\vspace{0.2cm}
\begin{center}
\setlength{\unitlength}{.0153in}
\begin{picture}(110,100)(30,500)

% Circles: quenched beta = 6.0. kappa_strange from K. Open circles used 
% for the 
% strange quark mass determined from K*.
% these are now commented out and kappa_strange from K taken as 
% circles for GPL plot. 
% Boxes: n_f = 2 beta = 5.6
% Error bars that are smaller than the symbols are not shown.
% axis
% update feb 1998  from final version fB paper      CTHD
\put(15,500){\line(0,1){100}}
\multiput(13,500)(0,50){3}{\line(1,0){4}}
\multiput(14,500)(0,10){11}{\line(1,0){2}}
\put(12,500){\makebox(0,0)[r]{{\large5.0}}}
\put(12,550){\makebox(0,0)[r]{{\large5.5}}}
\put(12,600){\makebox(0,0)[r]{{\large 6.0}}}
\put(12,570){\makebox(0,0)[r]{{\large GeV}}}
\put(15,500){\line(1,0){160}}

     \put(25,510){\makebox(0,0)[t]{{\large $B$}}}
     \put(26,528){\circle{6}}
     \multiput(20,527.9)(3,0){4}{\line(1,0){2}}
     \put(26,587.9){\circle{6}}
     \put(26,587.9){\line(0,1){8.3}}
     \put(26,587.9){\line(0,-1){8.3}}
     \put(38,599.5){\makebox(0,0)[t]{{\large $(2S)$}}}
     \multiput(20,586)(3,0){4}{\line(1,0){0.5}}
%     \put(27,583.7){$\!\!\Box$}
%     \put(27,585){\line(0,1){12}}
%     \put(27,585){\line(0,-1){12}}

     \put(55,510){\makebox(0,0)[t]{{\large $B^{*}$}}}
     \put(56,530.3){\circle{6}}
     \put(56,530.3){\line(0,1){0.5}}
     \put(56,530.3){\line(0,-1){0.5}}
     \multiput(50,532.6)(3,0){4}{\line(1,0){2}}
     \multiput(50,532.4)(3,0){4}{\line(1,0){2}}
%	\put(57,529.3){$\!\!\Box$}
%	\put(57,530.8){\line(0,1){.6}}
%	\put(57,530.8){\line(0,-1){.6}}
%	\put(57,591.8){$\!\!\Box$}
%	\put(57,593){\line(0,1){12}}
%	\put(57,593){\line(0,-1){12}}

     \put(80,510){\makebox(0,0)[t]{{\large $B_s$}}}
     \put(80,536.7){\circle{6}}
     \put(80,536.7){\line(0,1){2.9}}
     \put(80,536.7){\line(0,-1){1}}
%    \put(80,539.0){\circle{6}}
%     \put(80,539.0){\line(0,1){1.3}}
%     \put(80,539.0){\line(0,-1){1.3}}
     \multiput(75,538.1)(3,0){4}{\line(1,0){2}}
     \multiput(75,536.9)(3,0){4}{\line(1,0){2}}
     \put(80,592.8){\circle{6}}
     \put(80,592.8){\line(0,1){5.3}}
     \put(80,592.8){\line(0,-1){5.3}}
%    \put(80,593.7){\circle{6}}
%    \put(80,593.7){\line(0,1){4.8}}
%    \put(80,593.7){\line(0,-1){4.8}}
     \put(93,603.5){\makebox(0,0)[t]{{\large $(2S)$}}}
%    \put(83,538.0){$\!\!\Box$}
%     \put(84,539.4){\line(0,1){1}}
%     \put(84,539.4){\line(0,-1){1}}
%    \multiput(75,538.1)(3,0){4}{\line(1,0){2}}
%    \multiput(75,536.9)(3,0){4}{\line(1,0){2}}

     \put(105,510){\makebox(0,0)[t]{{\large $B^{*}_s$}}}
     \put(106,539.4){\circle{6}}
     \put(106,539.4){\line(0,1){2.9}}
     \put(106,539.4){\line(0,-1){1.0}}
%    \put(107,541.9){\circle{6}}
%     \put(107,541.9){\line(0,1){1.3}}
%     \put(107,541.9){\line(0,-1){1.3}}
     \multiput(100,542.8)(3,0){4}{\line(1,0){2}}
     \multiput(100,541.6)(3,0){4}{\line(1,0){2}}

     \put(147,510){\makebox(0,0)[t]{{\large $P-States$}}}
     \put(148,583.7){\circle{6}}
     \put(148,583.7){\line(0,1){5}}
     \put(148,583.7){\line(0,-1){5}}
     \put(161,588.8){\makebox(0,0){{\large $(B^*_2)$}}}
% The error on 3P2 comes from adding the statistical error (35 MeV) and the systematic 
% error from the difference between the 3P2T and the 3P2E (35 MeV) squared and taking 
% the square root thereof.
     \put(130,565.3){\circle{6}}
     \put(130,565.3){\line(0,1){3.8}}
     \put(130,565.3){\line(0,-1){3.8}}
     \put(139,561.5){\makebox(0,0)[t]{{\large $(B^*_0)$}}}
     \put(138,570.7){\circle{6}}
     \put(138,570.7){\line(0,1){3.4}}
     \put(138,570.7){\line(0,-1){3.4}}
%     \put(142,575.9){\circle*{3}}
%     \put(142,575.9){\line(0,1){3.4}}
%     \put(142,575.9){\line(0,-1){3.4}}
%    \put(145,576.1){$\!\!\Box$}
%    \put(145,577.5){\line(0,1){3.4}}
%    \put(145,577.5){\line(0,-1){3.4}}
     \put(128,579){\makebox(0,0){{\large $(B_1)$}}}
     \multiput(128,568.6)(3,0){8}{\line(1,0){2}}
     \multiput(128,571.0)(3,0){8}{\line(1,0){2}}
     \multiput(128,577.9)(3,0){8}{\line(1,0){0.5}}

\end{picture}
\end{center}
\caption{The $B$ meson spectrum. Dashed lines indicate experimental error 
bounds from the Particle Data Book, dotted lines, preliminary experimental 
results from DELPHI~\protect\cite{lep1,lep2}. }
\label{fig:meson_spect}
\end{figure}
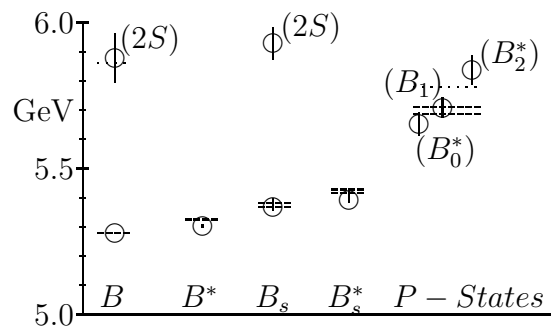

A summary of our results on the meson spectrum, compared with experimental
results, is given in Fig.~\ref{fig:meson_spect}. The error bars
include the statistical uncertainty, the error from interpolations and 
extrapolations to the physical quark masses, the statistical error in $a^{-1}$, 
and the uncertainty in fixing the strange quark mass.
The figure shows that  our calculation reproduces the presently known gross
features of the $B$ meson spectrum.
 
Now we discuss the fine and hyperfine structure in more detail.
\begin{figure}[t]
\vspace{-4cm}
\centerline{\hspace{1cm}
\epsfxsize=9.5cm
\epsfbox{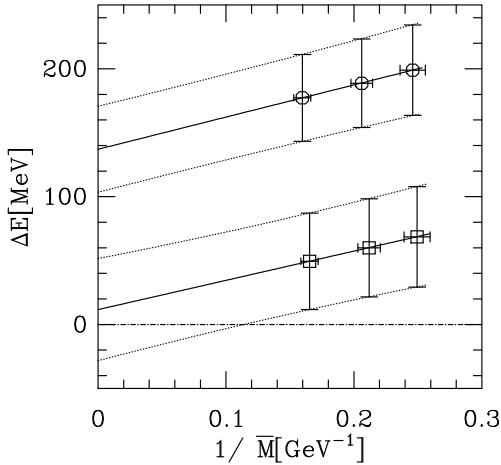}
}
\vspace{-2cm}
\caption{$B_2^\ast-B_0^\ast$ (circles) and $B_1-B_0^\ast$ (squares) splittings 
as a function of the spin-averaged heavy-light meson mass $\overline{M}$. 
Solid lines denote a linear fit in $1/\overline{M}$, dotted lines, the fit
errors. 
}
\label{fig:psplittings}
\end{figure}
For the $B^\ast-B$ splitting we obtain 24(5) MeV, which is significantly
smaller than the experimental value, 
45.8(4) MeV. Possible sources of this discrepancy are
quenching and discretization effects, corrections to the
perturbative coefficient of the $\vec{\sigma}\cdot\vec{B}$ term in the action,
and higher order  
relativistic corrections. For the $P$ state fine structure, the lattice provides
predictions as there are no experimental numbers available. In
heavy-light systems one differentiates $P$ states by the  light quark
angular momentum which can take the values $j_l =
1/2$ and $j_l = 3/2$. According to the coupling of the heavy quark spin, each
of these levels splits up further into a hyperfine doublet; for $j_l = 1/2$,
this consists of $B_0^\ast$ and $B_1$, and for $j_l = 3/2$, $B_1^\prime$ and
$B_2^\ast$. The $B_1$ and  the $B_1^\prime$ both have the quantum numbers $J^P 
= 1^+$ and can mix. Our two lattice operators for $J^P = 1^+$ give
two slightly different masses.  We assume that the lighter of our lattice
results  corresponds to the lighter physical state, which we expect to be the
$B_1$. We were however not able to separate the contributions of the physical
$1^+$ states to the heavier lattice state.

In the following we consider the $B_2^\ast-B_0^\ast$ and the $B_1-B_0^\ast$
splitting.  Both quantities, extrapolated to $\kappa_l$, are shown as a
function of the inverse heavy mass in Fig.~\ref{fig:psplittings}. According to 
our expectation, the $B_2^\ast-B_0^\ast$ is larger ($183(34)$ MeV) and has a
finite static limit. The $B_1-B_0^\ast$ splitting is $54(38)$ MeV,
comparable to the $B^\ast-B$ splitting, and its static limit is compatible
with zero.
\subsection{Baryons}
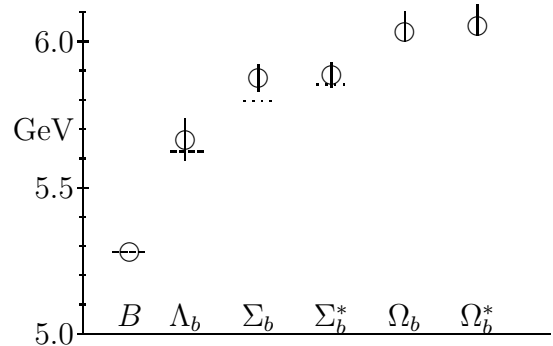
\begin{figure}
\vspace{0.8cm}
\begin{center}
\setlength{\unitlength}{.0153in}
\begin{picture}(110,100)(30,500)
% Circles: NRQCD quenched beta = 6.0 1/M^2 action, 
% Box: NRQCD n_f = 2 beta = 5.6, 
% Diamonds: clover c = 1 UKQCD Collaboration Phys.Rev. D54 (1996) 3619 quenched beta = 6.2 
% Triangle: C. Alexandrou et al, Phys. Lett. B337 (1994) 340-346, standard Wilson action
% extrapolated to the continuum from beta = 5.74, 6.0, 6.26.
% axis
% update feb 1998 with numbers from final version fB paper
\put(15,500){\line(0,1){110}}
\multiput(13,500)(0,50){3}{\line(1,0){4}}
\multiput(14,500)(0,10){12}{\line(1,0){2}}
\put(12,500){\makebox(0,0)[r]{{\large5.0}}}
\put(12,550){\makebox(0,0)[r]{{\large5.5}}}
\put(12,600){\makebox(0,0)[r]{{\large 6.0}}}
\put(12,570){\makebox(0,0)[r]{{\large GeV}}}
\put(15,500){\line(1,0){160}}

%\put(20,610){\makebox(0,0)[l]{\underline{\Large{{\bf Heavy-Light Baryons}}}}} 
%\put(20,620){\makebox(0,0)[l]{\Large{ ( Preliminary )  
%}}}

     \put(31,510){\makebox(0,0)[t]{{\large $B$}}}
     \put(31,528){\circle{6}}
     \multiput(25,527.9)(3,0){4}{\line(1,0){2}}

     \put(50,510){\makebox(0,0)[t]{{\large $\Lambda_b$}}}
     \put(50,566.5){\circle{6}}
     \put(50,566.5){\line(0,-1){7.1}}
     \put(50,566.5){\line(0,1){7.1}}
%     \multiput(39,561.5)(3,0){6}{\line(1,0){2}}
%     \multiput(39,563.3)(3,0){6}{\line(1,0){2}}
     \multiput(45,562.4)(3,0){4}{\line(1,0){2}}
%     \put(45.5,563){$\diamondsuit$}
%     \put(48,564){\line(0,1){6}}
%     \put(48,564){\line(0,-1){6}}
%     \put(50.3,571.8){$\triangle$}
%     \put(53,572.8){\line(0,1){14.5}}
%     \put(53,572.8){\line(0,-1){14.5}}
%     \put(43,581.8){$\!\!\Box$}
%     \put(43,583){\line(0,1){4}}
%     \put(43,583){\line(0,-1){4}}

     \put(75,510){\makebox(0,0)[t]{{\large $\Sigma_b$}}}
     \put(75,587.6){\circle{6}}
     \put(75,587.6){\line(0,-1){4.6}}
     \put(75,587.6){\line(0,1){4.6}}
%     \multiput(150,575.5)(3,0){4}{\line(1,0){2}}
%     \multiput(150,587.5)(3,0){4}{\line(1,0){2}}
     \multiput(70,579.7)(3,0){4}{\line(1,0){0.5}}
%     \put(100.5,576){$\diamondsuit$}
%     \put(103,577){\line(0,1){7}}
%     \put(103,577){\line(0,-1){7}}

     \put(100,510){\makebox(0,0)[t]{{\large $\Sigma_b^*$}}}
     \put(100,588.5){\circle{6}}
     \put(100,588.5){\line(0,1){4.3}}
     \put(100,588.5){\line(0,-1){4.3}}
%     \multiput(165,581.1)(3,0){4}{\line(1,0){2}}
%     \multiput(165,593.1)(3,0){4}{\line(1,0){2}}
     \multiput(95,585.3)(3,0){4}{\line(1,0){0.5}}
%     \put(150.5,577){$\diamondsuit$}
%     \put(153,578){\line(0,1){7}}
%     \put(153,578){\line(0,-1){7}}

     \put(125,510){\makebox(0,0)[t]{{\large $\Omega_b$}}}
     \put(125,603.4){\circle{6}}
     \put(125,603.4){\line(0,-1){3.3}}
     \put(125,603.4){\line(0,1){6.8}}

     \put(150,510){\makebox(0,0)[t]{{\large $\Omega^\ast_b$}}}
     \put(150,605.4){\circle{6}}
     \put(150,605.4){\line(0,-1){3.3}}
     \put(150,605.4){\line(0,1){7.1}}
\end{picture}
\end{center}
\vspace{-0.1cm}
\caption{Baryons with one $b$ and two light quarks. Dashed lines
indicate experimental error bounds from the Particle Data Book, dotted lines,
preliminary experimental results from DELPHI~\protect\cite{lep1}. 
}
\label{fig:bary_spect}
\vspace{-0.3cm}
\end{figure}
\begin{figure}[t]
\vspace{-4cm}
\centerline{
\hspace{1cm}
\epsfxsize=9.5cm
\epsfbox{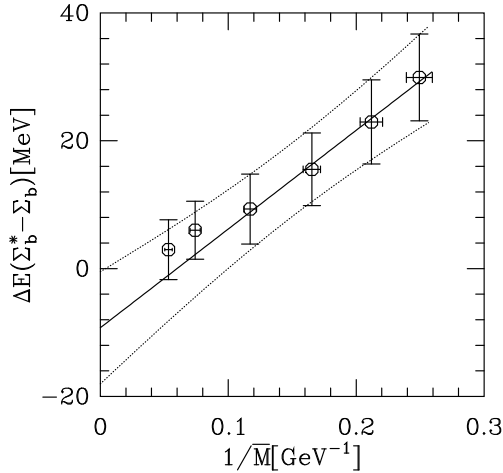}
}
\vspace{-1.8cm}
\caption{The $\Sigma^\ast_b-\Sigma_b$ splitting, extrapolated to $\kappa_l$,
as a function of the spin-averaged heavy-light meson mass $\overline{M}$. 
Solid lines
denote a linear fit in $1/\overline{M}$, dotted lines, the fit errors.
}
\vspace{-0.2cm}
\label{fig:Sigmastarsigma}
\end{figure}
In heavy-light baryons with one $b$ quark, the two light quarks can
couple to a sum spin $s_l = 0$ or $s_l = 1$. The former corresponds to the
$\Lambda_b$ baryon, whereas the configuration with $s_l = 1$ splits up into a
hyperfine doublet consisting of the $\Sigma_b$ and the $\Sigma^\ast_b$. 
An overview of the results for baryons with one $b$ and two light 
quarks is shown in Fig.~\ref{fig:bary_spect}. 
The errors are calculated the in same way as for Fig.~\ref{fig:meson_spect}.
The lattice results agree with 
experiment; the experimental values for the $\Sigma_b$ and
the $\Sigma^\ast_b$ are however still preliminary. 

The baryon hyperfine splitting $\Sigma^\ast_b-\Sigma_b$ as a function of the
inverse heavy mass is presented in Fig.~\ref{fig:Sigmastarsigma}. 
At the $b$ quark mass, we find $19(7)$ MeV. The extrapolation to infinite mass
is compatible with zero. 
\section{THE $B_c$ SPECTRUM}
To minimize quenching errors, one ideally fixes the lattice spacing from a
system that is sensitive to similar scales as the quantity that one is
extracting. Here, we restrict ourselves to a presentation of the data  quality 
and only a rough estimate of numbers, and do not perform a refined analysis. 
Since the $B_c$ is a heavy-heavy system, 
the charmonium $\overline{P}-\overline{S}$ splitting appears to be  
a reasonable choice. We use a preliminary 
number from Ref.~\cite{boyle} at $\beta=6.0$, $a^{-1} \sim 2.17$
GeV. 
%We then use a lattice $b$ mass parameter where for this lattice spacing
%the heavy-light meson mass agrees with the $B$ mass. $\kappa_{\rm
%charm}$ is fixed from the $\eta_c$  mass.
%At this lattice spacing,  
%the pseudoscalar heavy-light meson mass at $aM_0 = 2.0$ agrees with the
%physical $B$ meson mass. For clover quarks, the pseudoscalar heavy-heavy meson
%mass at $\kappa = 0.126$ agrees with the experimental $\eta_c$ mass. 
As $b$ and $c$ mass parameters for the $B_c$, we  use $aM_0 = 2.0$ and $\kappa =
0.126$. They correspond to physical $B$ and $\eta_c$ meson masses,
respectively, at this lattice spacing.

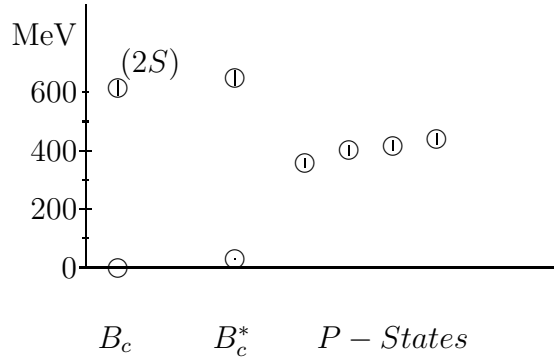
\begin{figure}[t]
\begin{center}
%\vspace{-1.5cm}
\setlength{\unitlength}{.0153in}
\begin{picture}(110,100)(30,100)

% Circles: quenched beta = 6.0. kappa_strange from K. Open circles used for the 
%strange quark mass determined from K*.
% these are now commented out and kappa_strange from K taken as 
% circles for GPL plot. 
% Boxes: n_f = 2 beta = 5.6
% Error bars that are smaller than the symbols are not shown.
% axis
% update feb 1998  from final version fB paper      CTHD
\put(15,100){\line(0,1){90}}
\multiput(13,100)(0,20){3}{\line(1,0){4}}
\multiput(14,100)(0,10){7}{\line(1,0){2}}
\put(12,100.0){\makebox(0,0)[r]{{\large 0}}}
\put(12,120.0){\makebox(0,0)[r]{{\large 200}}}
\put(12,140.0){\makebox(0,0)[r]{{\large 400}}}
\put(12,160.0){\makebox(0,0)[r]{{\large 600}}}
\put(12,180){\makebox(0,0)[r]{{\large MeV}}}
\put(15,100){\line(1,0){160}}

%\put(20,200){\makebox(0,0)[l]{\underline{\Large{{\bf $B_c$ Mesons 
%with NRQCD b-Quarks}}}}}
%\put(20,620){\makebox(0,0)[l]{\Large{ ( Preliminary )  
%}}}

     \put(25,80){\makebox(0,0)[t]{{\large $B_c$}}}
     \put(26,100){\circle{6}}
%     \multiput(20,527.9)(3,0){4}{\line(1,0){2}}
     \put(26,161.6){\circle{6}}
     \put(26,161.6){\line(0,1){2.5}}
     \put(26,161.6){\line(0,-1){2.5}}
     \put(37,175){\makebox(0,0)[t]{{\large $(2S)$}}}
%     \multiput(20,586)(3,0){4}{\line(1,0){0.5}}

     \put(65,80){\makebox(0,0)[t]{{\large $B_c^{*}$}}}
     \put(66,103){\circle{6}}
     \put(66,103){\line(0,1){0.1}}
     \put(66,103){\line(0,-1){0.1}}
%     \multiput(50,532.6)(3,0){4}{\line(1,0){2}}
%     \multiput(50,532.4)(3,0){4}{\line(1,0){2}}
     \put(66,164.9){\circle{6}}
     \put(66,164.9){\line(0,1){2.6}}
     \put(66,164.9){\line(0,-1){2.6}}

     \put(120,80){\makebox(0,0)[t]{{\large $P-States$}}}
     \put(135,144){\circle{6}}
     \put(135,144){\line(0,1){1.8}}
     \put(135,144){\line(0,-1){1.8}}
%     \put(135,144){\makebox(0,0){{\large $(B^*_2)$}}}
% The error on 3P2 comes from adding the statistical error (35 MeV) and the systematic 
% error from the difference between the 3P2T and the 3P2E (35 MeV) squared and taking 
% the square root thereof.
     \put(90,135.7){\circle{6}}
     \put(90,135.7){\line(0,1){1.6}}
     \put(90,135.7){\line(0,-1){1.6}}
%     \put(85,145){\makebox(0,0)[t]{{\large $(B^*_0)$}}}
     \put(105,140.1){\circle{6}}
     \put(105,140.1){\line(0,1){1.7}}
     \put(105,140.1){\line(0,-1){1.7}}
%     \put(110,135){\makebox(0,0){{\large $(B_1)$}}}
     \put(120,141.7){\circle{6}}
     \put(120,141.7){\line(0,1){1.6}}
     \put(120,141.7){\line(0,-1){1.6}}
%    \put(145,576.1){$\!\!\Box$}
%    \put(145,577.5){\line(0,1){3.4}}
%    \put(145,577.5){\line(0,-1){3.4}}
%     \put(120,140){\makebox(0,0){{\large $(B^\prime_1)$}}}
%     \multiput(128,568.6)(3,0){8}{\line(1,0){2}}
%     \multiput(128,571.0)(3,0){8}{\line(1,0){2}}
%     \multiput(128,577.9)(3,0){8}{\line(1,0){0.5}}

\end{picture}
\end{center}
\caption{$B_c$ level splittings. The error bars are purely statistical.}
\label{fig:Bcspect}
\end{figure}

A summary of the splittings between the excited states and the ground state is 
given in Fig.~\ref{fig:Bcspect}. 

The ground state mass is calculated in a similar way as for the $B$: 
\begin{equation}
M(B_c) = E_{\rm sim}(B_c) + \Delta_{\rm NRQCD} + \Delta_{\rm
clover}.\label{eq:bc} 
\end{equation}
The $\Delta$'s are calculated from heavy-heavy NRQCD or clover systems
respectively,  at the same bare $b$ and $c$ quark masses as used for the
$B_c$: 
\begin{equation}
\Delta = 1/2(M_{\rm kin}(HH) - E_{\rm sim}(HH)),.
\end{equation}
$E_{\rm sim}$ is the falloff of the respective meson correlator, and for
$M_{\rm kin}$ we use the definition of Ref.~\cite{fBpaper}. 
We obtain  6.3(1) GeV, whereas $M_{\rm kin}$ of the $B_c$ 
is 6.7(3) GeV. The difference is similar to the
statistical error in 
$M_{\rm kin}$, and is therefore not clearly significant. However it might be
an indication of discretization effects expected in heavy-heavy mesons
with clover quarks~\cite{andreas}. We are studying this further.
\subsubsection*{Acknowledgements}
This work has been supported by the U.S. DOE.
We thank the ACL at LANL and the NCSA at Urbana for
computational support.

\end{document}